\documentclass[
    ,final            
  ,psfig,graphix                 
  ]
  {aipproc}

\usepackage{epsfig}

\layoutstyle{6x9}

\def\beq{\begin{equation}}
\def\eeq{\end{equation}}

\def\bea{\arraycolsep .1em \begin{eqnarray}}
\def\eea{\end{eqnarray}}
\def\Tr{{\rm Tr}}

\def\eq#1{(\ref{#1})}

\def\s0#1#2{\mbox{\small{$ \frac{#1}{#2} $}}}
\def\0#1#2{\frac{#1}{#2}}

\begin{document}

\title{On fixed points of quantum gravity}

 \keywords {Quantum gravity, renormalisation group, fixed
  points, higher dimensions\\ {\bf Preprint:} SHEP-0542, CERN-PH/TH-2005-256}

\author{Daniel Litim}{
  address={
 School of Physics and Astronomy\\
University of Southampton, Southampton SO17
  1BJ, U.K.\\ and CERN, Theory Group, CH -- 1211 Geneva 23}
}

\begin{abstract}
  {We study the short distance behaviour of euclidean quantum gravity in the
    light of Weinberg's asymptotic safety scenario. Implications of a
    non-trivial ultraviolet fixed point are reviewed. Based on an optimised
    renormalisation group, we provide analytical flow equations in the
    Einstein-Hilbert truncation. A non-trivial ultraviolet fixed point is
    found for arbitrary dimension. We discuss a bifurcation pattern in the
    spectrum of eigenvalues at criticality, and the large dimensional limit of
    quantum gravity. Implications for quantum gravity in higher dimensions are
    indicated.}
\end{abstract}

\maketitle


\section{Introduction}

Gravitational interactions at distances sufficiently large compared to the
Planck length are described by the classical theory of general relativity. At
smaller length scales, quantum effects are expected to become important. The
quantisation of general relativity, however, still poses problems. It is known
since long that four-dimensional quantum gravity is perturbatively
non-renormalisable, meaning that an infinite number of parameters have to be
fixed to renormalise standard perturbation theory.  One may wonder whether a
quantum theory of gravity in terms of the metric degrees of freedom can exist
as a well-defined, non-trivial and cutoff-independent local theory down to
arbitrarily small distances. It is generally believed that the above
requirements imply the existence of a non-trivial ultraviolet (UV) fixed point
under the renormalisation group (RG), governing the short-distance physics.
Then it would suffice to adjust a finite number of parameters, ideally taken
from experiment, to make the theory asymptotically safe. The corresponding
short distance fixed point action would then provide a valid microscopic
starting point to access classical general relativity as a ``low energy
phenomenon'' of a local quantum field theory in the metric field.

For quantum gravity, this asymptotic safety scenario has been introduced by
Weinberg \cite{Weinberg}. In the vicinity of two dimensions, a non-trivial
fixed point has been identified within perturbation theory, to leading
\cite{Weinberg}-\cite{Christensen:1978sc} and subleading order
\cite{Aida:1996zn} in $\epsilon=d-2\ll 1$.  In the last couple of years,
non-perturbative renormalisation group studies have been performed in the
four-dimensional case. A non-trivial fixed point has been detected within
various renormalisation group studies, also including higher dimensional
operators or non-interacting scalar, vector and matter fields
\cite{Reuter:1996cp}- \cite{Peter}. Additional indications for the existence
of a non-trivial fixed point in four dimensions have been provided through
lattice simulations within both Regge's simplicial lattice formulation
\cite{Hamber:1999nu} and the causal dynamical tringulations approach,
$e.g.$~\cite{Ambjorn:2005qt}.

From a renormalisation group point of view, the ``critical'' dimension of
quantum gravity is $d_{\rm cr} =2$.\footnote{This is different from most
  particle physics theories, where the critical dimension of the relevant
  couplings is $d=4$, $e.g.$~QED, QCD.} In two dimensions, the gravitational
coupling has vanishing canonical dimension, and standard perturbation theory
is applicable. In turn, for any $d>d_{\rm cr}$ the gravitational coupling has
negative mass dimension, indicating that the theory is perturbatively
non-renormalisable. Hence, one expects that the local renormalisation group
properties of quantum theories of gravity for different dimensions with
$d>d_{\rm cr}$ share qualitative properties. We conclude that the case of four
dimensions, from a quantum gravity point of view, is by no means
distinguished.  Continuity in the dimension, together with the indications for
a non-trivial fixed point in four dimensions, suggest that a non-trivial fixed
point should persist, to the least, in some vicinity of four dimensions. We
emphasize that this heuristic line of reasoning is solely based on local RG
properties of the theory, and insensitive against global properties within
specific dimensions.

In this contribution, we discuss the asymptotic safety scenario in the context
of quantum gravity. Based on a wilsonian renormalisation group, we provide
unique analytical fixed point solution in the Einstein-Hilbert truncation for
any dimensions $d>d_{\rm cr}$ \cite{Litim:2003vp}. The approach is related to
the integrating-out of momentum modes from a path integral representation of
the theory \cite{ERG-Reviews}, ammended by an appropriate optimisation
\cite{Litim:2000ci,Litim:2001up}. Consequently, the reliability of results
based on optimised flows is enhanced \cite{Litim:2002cf}. Results and
implications for quantum gravity in higher dimensions are discussed.

\section{Asymptotic safety}

We recall a few general requirements and implications of the asymptotic safety
scenario and a non-trivial fixed point in quantum gravity \cite{Weinberg}.
First of all, the asymptotic safety scenario relies on the existence of a
(non-trivial) fixed point at short distances.  This generalises a pattern
observed for perturbatively renormalisable theories, which often are related
to a non-interacting UV fixed point, $e.g.$~asymptotic freedom
in QCD.  Secondly, it is mandatory that the short-distance fixed point is
connected with the long-distance behaviour of the theory by a well-defined
renormalisation group trajectory. Elsewise the putative fixed point would
remain disconnected from the known physics at large distances. Finally, it is
required that the UV fixed point displays at most a finite number of
(infrared) unstable directions. Elsewise, the predictive power is spoiled,
because an infinite number of unstable directions would require the
fine-tuning of infinitely many parameters in order to reach the IR limit.
Then, the fixed point together with the RG trajectory serve as a fundamental
definition of the theory.

We proceed with a discussion of the renormalisation group flow for the
gravitational coupling $G$ in $d$ dimensions. We introduce the renormalised
dimensionless coupling as $g=\mu^{d-2}\,Z^{-1}_G(\mu)\, G$ and the graviton
wave function renormalisation factor $Z_G(\mu)$, normalised as $Z_G(\mu_0)=1$
at $\mu=\mu_0$. The momentum scale $\mu$ denotes the renormalisation scale,
and can be thought of as an energy scale $E$, or momentum transfer $p$, or,
more formally, as some wilsonian momentum cutoff $k$ as used below. The
graviton anomalous dimension, given by $\eta=-\mu\partial_\mu \ln Z_G$, is a
non-trivial function of $g$ and other couplings parametrising the theory. Then
the RG flow reads
\beq\label{beta-g}
\mu\partial_\mu g = (d-2+\eta)\,g\,.
\eeq
From \eq{beta-g} we conclude
that the gravitational coupling displays a non-interacting (gaussian) fixed
point $g_*=0$. This entails $\eta=0$, $i.e.$ classical scaling. On the other
hand, non-trivial RG fixed points, if they exist, correspond to the implicit
solutions of
\beq\label{eta}
\eta_*=2-d\,.
\eeq
Hence, a non-trivial fixed point of quantum gravity in $d>2$ implies a
negative integer value for the graviton anomalous dimension, precisely
counter-balancing the canonical dimension of $G$. 

Integer values for anomalous dimensions are well-known from other gauge
theories at criticality: in the $d$-dimensional abelian higgs theory, for
example, the abelian charge $e^2$ has mass dimension $[e^2]=4-d$, whence
$\beta_{e^2}=(d-4+\eta)\, e^2$. In three dimensions, the theory displays a
non-perturbative infrared fixed point at $e^2_*\neq 0$, leading to $\eta=1$
\cite{Bergerhoff:1995zq}. The fixed point belongs to the universality class of
standard superconductors with the charged scalar field describing the Cooper
pair. The integer value $\eta =1$ implies that the magnetic field penetration
depth and the Cooper pair correlation length scale with the same universal
critical exponent at the phase transition
\cite{Bergerhoff:1995zq,Herbut:1996ut}. In turn, scalar theories at
criticality, not protected by a local gauge symmetry, often display
non-integer anomalous dimensions characterising the universality class.

In quantum gravity, a non-trivial fixed point behaviour leads to two important
implications. First of all, the dimensionful gravitational coupling constant
scales as $G(\mu)\to g_*/\mu^{d-2}$ at the fixed point. In the case of an
ultraviolet fixed point for diverging of $\mu$, the coupling $G$ becomes
arbitrarily small. This indicates that gravity might become asymptotically
free at short distances, similar to QCD. Conversely, in case of an infrared
fixed point for vanishing $\mu$, the dimensionful coupling grows large. This
behaviour implies non-trivial long distance modifications of gravity,
$e.g.$~\cite{Bentivegna:2003rr}. Secondly, the scalar part of the renormalised
graviton propagator scales $\sim p^{-2+\eta}$.  Here we have identified $\mu$
with the momentum scale $p$. At an UV fixed point, this leads to an increased
momentum decay $\sim p^{-d}$, which, in position space, corresponds to a
logarithmic behaviour $\sim \ln (|x-y|\mu)$ of the propagator, reminiscent
from two dimensions.  Hence, the integer value of \eq{eta}, on the level of
the graviton propagator, implies a dimensional crossover from $d$-dimensional
behaviour in the perturbative regime to an effectively two-dimensional
behaviour in the vicinity of a non-trivial fixed point. A similar result is
found for the spectral dimension of quantum gravity space-times at short
distances \cite{Lauscher:2005qz,Ambjorn:2005qt}.

\section{Renormalisation group}

Whether or not the non-trivial fixed point is realised in quantum gravity can
only be assessed by an explicit renormalisation group study.  We apply a
wilsonian renormalisation group, based on a momentum cutoff for the propagting
degrees of freedom (for reviews, see \cite{ERG-Reviews}). It is based on a
scale-dependent action functional $\Gamma_k$ of the mean gravitational field
$\langle g_{\mu\nu}\rangle_k$, where $k$ denotes a wilsonian momentum cutoff
scale.  The wilsonian flow equation describes the change of $\Gamma_k$ under
an infinitesimal variation of $k$.  By construction, $\Gamma_k$ comprises all
quantum fluctuations down to the momentum scale $q^2\approx k^2$. The flow
interpolates between some microscopic action at short distances $k\to\infty$
and the full quantum effective action at large distances $k\to 0$.
Diffeomorphism invariance under local coordinate transformations is controlled
by modified Ward identities \cite{Reuter:1996cp}, similar to those known for
non-Abelian gauge theories \cite{Freire:2000bq}.  In its modern formulation,
the flow with respect to the logarithmic scale parameter $t=\ln k$ is given by
\beq\label{ERG} 
\partial_t \Gamma_k=
\frac{1}{2} \Tr \frac{1}{\Gamma_k^{(2)}+R_k}\partial_t R_k\,.  
\eeq 
Here, the trace stands for a momentum integration and a sum over indices and
fields, and $R_k$ denotes an appropriate wilsonian momentum cutoff at momentum
scale $q^2\approx k^2$. For quantum gravity, we consider the flow \eq{ERG} for
$\Gamma_k$ in the Einstein-Hilbert truncation
\begin{equation}\label{EHk}
 \Gamma_k=
\0{1}{16\pi G_k}\int d^dx \sqrt{g}\left[-R(g)+2\bar\lambda_k\right]
\ + \ {\rm classical\ gauge\ fixing}\,,
\end{equation}
retaining the volume element and the Ricci scalar as independent operators. In
\eq{EHk} we have introduced the gravitational coupling constant $G_k$ and the
cosmological constant $\bar \lambda_k$.  The Ansatz \eq{EHk} differs from the
standard Einstein-Hilbert action in $d$ Euclidean dimensions by the fact that
the gravitational coupling and the cosmological constant have turned into
running couplings.  We introduce dimensionless renormalised gravitational and
cosmological constants $g_k$ and $\lambda_k$ as
\begin{equation}\label{glk}
g_k=k^{d-2}\, G_k\, \equiv k^{d-2}\, Z^{-1}_{N,k}\ \bar G\,,
\quad\quad
\lambda_k=\,k^{-2}\, \bar\lambda_k\,.
\end{equation}
and $Z_{N,k}$ denotes the wave function renormalisation factor for the
Newtonian coupling. Their flows follow from \eq{ERG} by an appropriate
projection onto the operators in \eq{EHk}. 
For explicit constructions of a momentum cutoff in the background field gauge
with gauge fixing parameter $\alpha$, see
\cite{Reuter:1996cp,Lauscher:2001ya}. Closely related flows have also been
derived in \cite{Bonanno:2004sy} within a proper-time approximation
\cite{Litim:2001hk}. Here, we use the approach of
\cite{Reuter:1996cp,Lauscher:2001ya} for the tensorial structure of the
momentum cutoff in addition with an optimised momentum cutoff
\cite{Litim:2001up} for its scalar part.  This leads to an analytical flow
equation
\bea
\label{beta-l-inf}
\partial_t\lambda&\equiv&\beta_{\lambda}=
\frac{P_1}{
P_2+4(d+2) g}\quad\quad
\partial_t g\equiv\beta_{ g}
=\frac{(d-2)\, g\, P_2}{
P_2+4(d+2) g}\\
\label{P1}
P_1&=&
-16 \lambda^3 
+ 4  \lambda^2 (4 - 10 d\,  g - 3 d^2\,  g + d^3\, g) 
+4\lambda (10 d \, g +  d^2\,  g -  d^3 \, g-1) 
\nonumber\\ &&
+ d (2 + d) (d - 16\,  g + 8 d\,  g-3) \, g 
\\
\label{P2}
P_2&=& 8 ( \lambda^2 - \lambda -d\, g)+ 2 \,.  \eea For convenience, a factor
$1/\alpha$ is absorbed into the definition \eq{glk}, and a factor
$c_d=(4\pi)^{d/2-1}\Gamma(d/2+2)$ is absorbed into the definition of $g$.
Furthermore, we have performed the limit $\alpha\to\infty$ in \eq{beta-l-inf},
where the results take their simplest form. The flow for arbitrary $\alpha$ is
given in \cite{Litim2005}. The flow of $ g$ vanishes identically in two
dimensions. For the anomalous dimension, we find
\bea\label{eta-inf}
\eta&=&     
\frac{(d+2)\, g}{ g- g_{\rm bound}}\,,\quad\quad\quad
g_{\rm bound}(\lambda)=
\frac{\left(1-2\lambda\right)^2}{2(d-2)}\,.
\eea
The anomalous dimension diverges at $g=g_{\rm bound}(\lambda)$, which limits
the domain of validity. For all real $\lambda$ and $d>2$, we have $g_{\rm
  bound}\ge 0$. The anomalous dimension vanishes for $ g=0$, and in $d=2$
dimensions.

\section{Analytical fixed points}

Next we identify the non-trivial fixed points of $\beta_{ g}(\lambda_*,
g_*)=0=\beta_{ \lambda}(\lambda_*, g_*)$.  From $P_2=0$, \eq{P2}, we deduce
that $g_*=g_{\rm bound}(\lambda_*)\times(d-2)/(2d)$. With \eq{eta-inf},
we conclude that the gravitational coupling fixed point is positive for any
real fixed point $\lambda_*$. Inserting this result into \eq{P1}, equating
$P_1=0$, and factoring-out a common factor $(1-2\lambda)^2$ leads to a simple
quadratic equation with two real solutions, as long as $d\ge
(1+\sqrt{17})/2\approx2.56$.  The physical fixed point obeys $\lambda<\s012$
and reads
\bea
\label{sol-inf}
\lambda_*&=&
\frac{d^2-d-4-\sqrt{2d(d^2-d-4)}}{2(d-4)(d+1)}\,,
\quad
g_*=
\frac{\Gamma(d/2+2)}{(4\pi)^{1-d/2}}
\frac{(\sqrt{d^2-d-4}-\sqrt{2d})^2}{2(d-4)^2(d+1)^2}\,.\quad
\eea
Here, we have reinserted the factor $c_d$ in $g$. The solution \eq{sol-inf} is
continuous and well-defined for all $d\ge 2.56$ and becomes complex for lower
dimensions. For general $\alpha<\infty$, the fixed point extends down to $d=2$.

Fixed points are non-universal quantities and may depend on unphysical
parameters. In turn, the rates at which small perturbations about the fixed
point grow with scale are universal. These are denoted as $-\theta$, or
$-1/\nu$ in the statistical physics literature, and given by the eigenvalues
of the stability matrix at criticality. In four dimensions, and reinserting
$c_d$, we find $\theta=\theta'+i\theta''$, with
\bea && 
\lambda_*=\frac{1}{4}              \,,\quad 
g_*=\frac{3\pi}{8}                 \,,\quad
\theta'=\frac{5}{3}                \,,\quad 
\theta''=\frac{\sqrt{167}}{3}      \,,\quad
|\theta|=\frac{8}{\sqrt{3}}        \,.  
\eea 
This compares well with the lattice result $\nu\approx 1/3$ of
\cite{Hamber:1999nu}. In Fig.~1, the trajectory connecting the UV fixed point
with the gaussian one is given, together with the running couplings along the
separatrix. The complex eigenvalue is reflected by a rotating trajectory at
the UV fixed point. It originates from a strong mixing of the scaling of the
volume operator and the Ricci scalar. Along the separatrix, the
 anomalous dimension displays a cross-over from a perturbatively small value towards the
non-trivial fixed point value \eq{eta}. The fixed point and the flow pattern
persists for arbitrary gauge fixing parameter.

\begin{figure}[t]
\includegraphics[width=.5\textwidth]{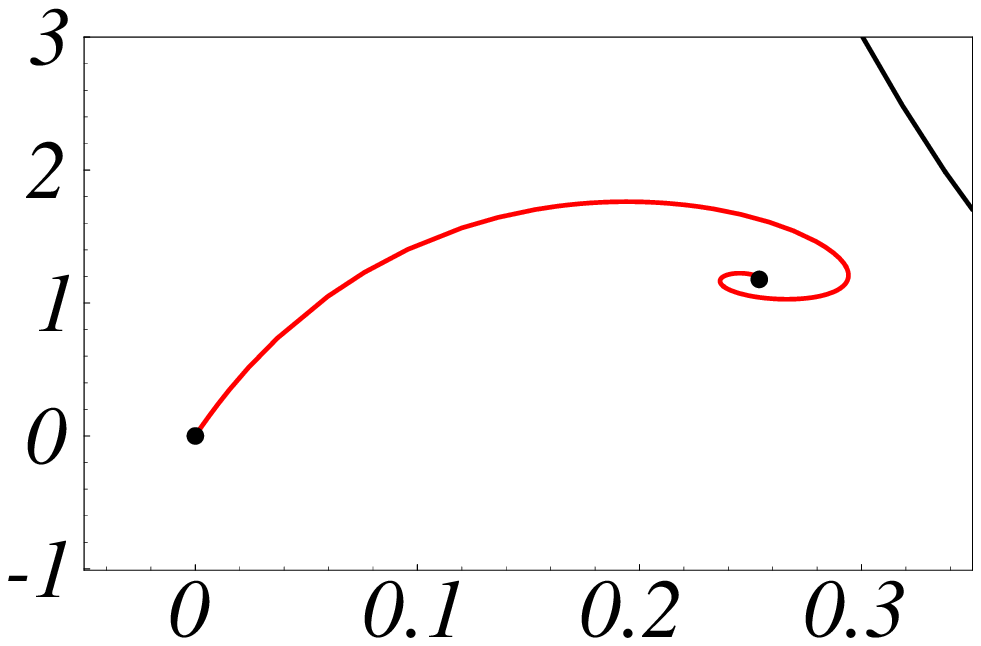}
\includegraphics[width=.5\textwidth]{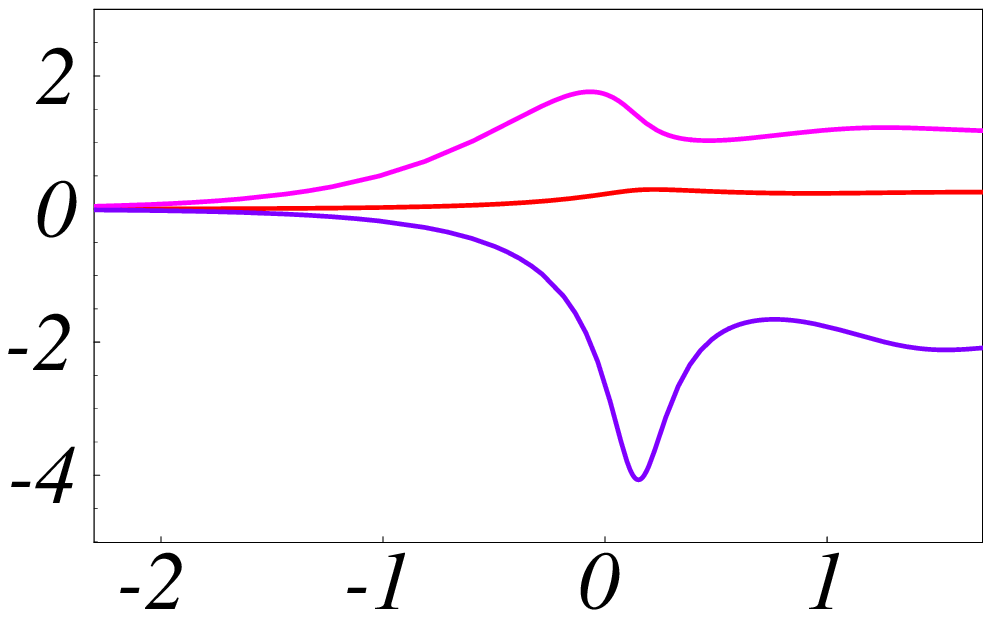}
\caption{Phase diagram of quantum gravity in four dimensions.
  Left panel: The separatrix $g(\lambda)$ in four dimensions (red line),
  connecting the non-trivial UV fixed point (right dot) with the gaussian
  fixed point (left dot). The thin full line in the upper right corner denotes
  $g=g_{\rm bound}$, see \eq{eta-inf}. Right panel: running couplings along
  the separatrix.  From top to bottom: the dimensionless gravitational
  coupling (violet line), the dimensionless cosmological constant (red line)
  and the anomalous dimension (blue line) as a function of $\ln (k/M_{\rm
    Pl})$ in four dimensions. The strong correlation between the scaling of
  the volume element $\sqrt{g}$ and the Ricci scalar $\sqrt{g}R$ in the
  vicinity of the fixed point is responsible for the rotation of the
  separatrix, and for the oscillatory behaviour of the trajectories at scales
  $k> M_{\rm Pl}$.}
\end{figure}

\section{Higher dimensions}

The fixed point \eq{sol-inf}, and the corresponding solution for arbitrary
gauge fixing parameter \cite{Litim:2003vp}, is found independently of the
dimension. The fixed point is unique, real, positive, and continuously
connected to the perturbatively known fixed point in two dimensions. This is
quite remarkable, also in view of the fact that the Einstein-Hilbert
truncation is expected to be more sensitive to higher dimensional operators in
higher than in lower dimensions. For a detailed study of the general cutoff
and gauge fixing independence of this fixed point, see~\cite{Peter}.

Here, we point out three noteworthy aspects of the fixed point for general
dimension. For more details, see \cite{Litim2005}.  First of all, we notice a
non-trivial bifurcation in the eigenvalue spectrum, as a function of
dimension.  For $2<d_{\rm low}<d<d_{\rm up}<\infty$, the universal eigenvalues
of the stability matrix are complex, and real otherwise (see
Tab.~\ref{tab-bifurcation} for numerical values). Complex eigenvalues indicate
that the operators $\sqrt{g}$ and $\sqrt{g}R$ scale with a similar strength,
and that they are subject to strong mixing effects, parametrised by the
off-diagonal elements of the Jacobi matrix.  This is the case in four
dimensions. In turn, real eigenvalues indicate that the scaling behaviour at
the fixed point is dominated by the volume element, while the Ricci scalar
remains subleading. This behaviour is well-known in the vicinity of two
dimensions. The new result is that a similar behaviour persists
for sufficiently large dimension. The large dimensional limit, in this light,
shares an important similarity with the limit of small dimensions. If this
structure persists in the full theory, we expect that an expansion about two
dimensions (in inverse dimension) has a finite radius of convergence, given by
the lower (upper) critical dimension. 

Secondly, it is possible to perform the large dimensional limit. For any
$\alpha\in[0,1]$ (and similarily for other $\alpha$), one finds the universal
eigenvalues $\theta_1=d^3/156$ and $\theta_2=24d/13$, related to the
scaling of the $\sqrt{g}$ and $\sqrt{g}R$ operator, respectively.  The index
$\nu=1/\theta_2$ agrees very well with the result $\nu=1/(2d)$ as obtained
previously for vanishing cosmological constant $\lambda=0$
\cite{Litim:2003vp}. This should also be compared with $\nu=1/(d-1)$ based on
geometrical considerations \cite{Hamber:2004ew}, with $\nu\approx 1.9/d$ based
on an extrapolation of low dimensional lattice results \cite{Hamber:2005vc},
and $\nu=1/(d-2)$ as obtained within a perturbative expansion about two
dimensions \cite{Gastmans:1977ad,Christensen:1978sc}.

Finally, a non-trivial fixed point of quantum gravity in higher dimensions is
of immediate interest for phenomenological applications. It has been suggested
that the fundamental Planck scale may be as low as the electroweak scale, if
gravity propagates in a $d=4+n$ dimensional ``bulk'' with $n$ ``extra''
spatial dimensions, while standard model particles only propagate in a
four-dimensional ``brane''\cite{Arkani-Hamed:1998rs,Antoniadis:1990ew}. Then high-energy particle
physics experiments at LHC are potentially sensitive to the fundamental Planck
scale and quantum gravitational effects. Up to now, phenomenological effects
due to higher dimensional gravity have been studied using classical
propagators and vertices, ammended by an ultraviolet cutoff,
$e.g.$~\cite{Giudice:1998ck}. In turn, a fixed point scenario as described
here is ultraviolet finite and would not require an additional UV
regularisation. It will be interesting to identify physical observables most
strongly sensitive to the above scenario.

\begin{table}
\begin{tabular}{c|cccccc}
\hline
\tablehead{1}{r}{b}{$\alpha$}&
$\quad\quad{}     0            \quad\quad{}    $&
$\quad\quad{}     \s012        \quad\quad{}    $&
$\quad\quad{}     1            \quad\quad{}    $&
$\quad\quad{}     1_{\rm A}    \quad\quad{}    $&
$\quad\quad{}     2            \quad\quad{}    $&
$\quad\quad{}     \infty       \quad\quad{}    $
\\[.5ex] \hline 
\tablehead{1}{r}{b}{$    d_{\rm low}   $}&
$\ \ 2.900$& 
$\ \ 2.901 $& 
$\ \ 2.872$&
$\ \ 2.756 $&
$\ \ 2.765 $&
$\ \ 2.562 $
\\
\tablehead{1}{r}{b}{$    d_{\rm up}    $}&
$23.727$& 
$23.672 $&
$24.282$&
$23.985$& 
$17.394$&
$21.381$\\
\hline
\end{tabular}
\caption{\label{tab-bifurcation} Lower and upper critical
    dimensions as functions of the gauge fixing parameter. The data is
    obtained using a wilsonian momentum cutoff with tensorial momentum 
    structure as in \cite{Reuter:1996cp} (for $\alpha=1_{\rm A}$) or
    \cite{Lauscher:2001ya} (for general $\alpha\ge 0$), and a scalar
    momentum structure as in 
    \cite{Litim:2001up}.   }
\end{table}

\section{Conclusions}
We have discussed basic implications of an asymptotic safety scenario for
quantum gravity. It has been emphasized that the case of four space-time
dimensions, from a renormalisation group point of view, is not distinguished.
This is reflected by the non-trivial fixed point structure  in the Einstein
Hilbert theory, which displays an UV fixed point for arbitrary dimension,
$e.g.$~\eq{sol-inf}. Maximal reliability in the present truncation is
guaranteed by the underlying optimisation. It was pointed out that the large
dimensional limit -- in view of its scaling properties at criticality -- shares
qualitative similarities with the low dimensional limit, the vicinity of
$d=2$. In contrast, in the neighbourhood of four dimensions, scaling at
criticality is characterised by strong operator mixing. This has lead to an
interesting bifurcation pattern in the eigenvalue spectrum.

In four dimensions and below, our results are fully consistent with previous
renormalisation group studies. The fixed point is remarkably stable, with only
a mild dependence on the gauge fixing parameter. The phase diagram is equally
robust. Furthermore, it is encouraging that the qualitative picture achieved
so far is backed-up by lattice simulations.  The fixed point consistently
extends to higher dimensions, a region which previously has not been
accessible. Hence it is likely that the fixed point exists in the full theory.
We expect that the analytical form of the flow, crucial for the present
analysis, is equally useful in extended truncations. If the above picture
persists in these cases, quantum gravity exists as a well-defined local
quantum field theory in the metric field down to arbitrarily short distances.


\begin{theacknowledgments}
This work is supported by an EPSRC Advanced Fellowship.  
\end{theacknowledgments}

\end{document}